\documentclass[11pt]{article}
\usepackage{graphicx}
\usepackage{amssymb}
\usepackage{epstopdf}
\usepackage{natbib}

\newcommand{\V}{{\rm V}}
\newcommand{\E}{{\rm E}}
\renewcommand{\P}{{\rm P}}

\begin{document}

\title{A preferential attachment model with Poisson growth for scale-free networks}

\author{Paul Sheridan \quad Yuichi Yagahara \quad Hidetoshi Shimodaira\\
\\
              Tokyo Institute of Technology, \\
              Department of Mathematical and Computing Sciences, \\
              2-12-1 Ookayama, Meguro-ku, Tokyo 152-8552, Japan \\
              \tt{sherida6@is.titech.ac.jp}             \\
}

%\institute{Paul Sheridan (corresponding author) \and Yuichi Yagahara \and Hidetoshi Shimodaira \at
%              Tokyo Institute of Technology, \\
%              Department of Mathematical and Computing Sciences, \\
%              2-12-1 Ookayama, Meguro-ku, Tokyo 152-8552, Japan \\
%              \email{sherida6@is.titech.ac.jp}           %  \\
%}

\maketitle

\begin{abstract}
We propose a scale-free network model with a tunable power-law exponent. The Poisson growth model, as we call it, is an offshoot of the celebrated model of Barab\'{a}si and Albert where a network is generated iteratively from a small seed network; at each step a node is added together with a number of incident edges preferentially attached to nodes already in the network. A key feature of our model is that the number of edges added at each step is a random variable with Poisson distribution, and, unlike the Barab\'{a}si-Albert model where this quantity is fixed, it can generate any network. Our model is motivated by an application in Bayesian inference implemented as Markov chain Monte Carlo to estimate a network; for this purpose, we also give a formula for the probability of a network under our model.

%\keywords{Bayesian inference \and complex networks \and  network models \and power-law \and scale-free}
\end{abstract}

\vspace{2em}


\section{Introduction} \label{sec: intro}

Until recent times, modeling of large-scale, real-world networks was primarily limited in scope to the theory of {\it random networks} made popular by~\cite{Erdos}. In the Erd\"{o}s-R\'{e}nyi model, for instance, a network of $N$ nodes is generated by connecting each pair of nodes with a specified probability. The {\it degree distribution} $p(k)$ of a large-scale random network is described by a binomial distribution, where the {\it degree} $k$ of a node denotes the number of undirected edges incident upon it. Thus, degree in a random network has a strong central tendency and is subject to exponential decay so that the average degree of a network is representative of the degree of a typical node.

Over the past decade, though, numerous empirical studies of {\it complex networks}, as they are known, have established that in many such systems---networks arising from real-world phenomena as diverse in origin as man-made networks like the World Wide Web, to naturally occurring ones like protein-protein interaction networks, to citation networks in the scientific literature; see~\cite{Albert2}, \cite{Jeong}, and \cite{Redner}, respectively---the majority of nodes have only a few edges, while some nodes, often called {\it hubs}, are highly connected. This characteristic cannot be explained by the theory of random networks. Instead, many complex networks exhibit a degree distribution that closely follows a {\it power-law} $p(k) \propto k^{-\gamma}$ over a large range of $k$, with an exponent $\gamma$ typically between 2 and 3. A network that is described by a power-law is called {\it scale-free}, and this property is thought to be fundamental to the organization of many complex systems;~\cite{Strogatz2}.

As preliminary experimental evidence mounted (see \cite{Watts}, for instance), a simple, theoretical explanation accounting for the universality of power-laws soon followed; the network model of~\cite{Barabasi} (BA) provided a fundamental understanding of the development of a wide variety of complex networks on an abstract level. Beginning with a connected seed network of $t_0$ nodes, the BA algorithm generates a network using two basic mechanisms: {\it growth}, where over a series of iterations $t = t_0, t_0+1, \ldots$ the network is augmented by a single node together with $m \leq t_0$ undirected incident edges; and {\it preferential attachment} where the new edges are connected with exactly $m$ nodes already in the network such that the probability a node of degree $k$ gets an edge is proportional to $r(k)=k$, the degree of the node. When $m$ is fixed throughout, \cite{Bollobas} showed rigorously that the BA model follows a power-law with exponent $\gamma=3$ in the limit of large $t$. The sudden appearance of the BA model in the literature, nearly a decade ago, sparked a flurry of research in the field, and, consequently, numerous variations and generalizations upon this prototypal model have been proposed;~\cite{Albert} and~\cite{Newman} rank as the preeminent survey papers on the subject. 

In this paper, we propose a new {\it growing model} based on preferential attachment: the Poisson growth (PG) model. Our model, as described in Section~\ref{sec: PG model}, is an extension of the BA model in two regards. Firstly, we consider the number of edges added at a step to be a random quantity; at each step, we assign a value to $m$ according to a Poisson distribution with expectation $\lambda > 0$. Secondly, we avail ourselves of a more general class of preferential attachment functions $r(k)$ studied by several authors including~\cite{Krapivsky} and~\cite{Dorogovtsev}. In Section~\ref{sec: degree distribution} we argue that the degree distribution of the PG model follows a power-law with exponent $\gamma$ that can be tuned to any value greater than 2; the technical details of our argument are left for the Appendix. In addition, we conducted a simulation study to support our theoretical claims. Our results, provided in Section~\ref{sec: simulation}, show that the values of $\gamma$ we estimated from networks generate under the PG model are in agreement with those predicted by our formulae for the power-law exponent.

Our motivation for proposing the PG model, as explained in Section~\ref{sec: discussion}, arises from a need for a simple, yet realistic model that is serviceable in applications. In fact, with our model every network has a nonzero probability of being generated, in addition to possessing a tunable power-law exponent. In contrast, the BA model has a fixed $\gamma$, and is subject to numerous structural constraints which severely limit the variety of generable networks. We give a simple formula for the probability of a network under the PG model, which can be applied quite naturally in Bayesian inference using Markov chain Monte Carlo (MCMC) methods. Firstly, given a network $G$, we may estimate the PG model parameters, or engage in model selection in the case when we have more models; or, going against the grain, we may estimate an unknown $G$ from data using our PG model formula as a scale-free prior distribution.

Finally, other scale-free models have been put forth in the literature, besides our new model, that are realistic enough for use in applications. An extension of BA model by~\cite{Albert3} incorporates ``local events,'' which allows for modifications, such as rewiring of existing edges, to the network at each step. Other authors, including~\cite{Sole}, proposed a class of growing models based on node duplication and edge rewiring. Other scale-free model not based on growth have also been proposed; for example, the static model of~\cite{Lee}. Among these models, the PG model is the simplest preferential attachment model sufficiently realistic for use in applications.

\section{The Poisson Growth Model} \label{sec: PG model}

In the PG model, we begin with a small seed network of $t_0$ nodes. Let $G_t=(V_t,E_t)$ be the network at the onset of time step $t \geq t_0$ where $V_{t}=\{v_1, v_2, \ldots, v_{t} \}$ is a set of $t$ nodes and $E_t$ is a multiset of undirected edges so that multiple edges between nodes, but no loops, are permitted. The updated network $G_{t+1}$ is generated from $G_t$ as follows: 
\begin{description}
	\item[{\it Poisson growth}:] A new node $v_{t+1}$ is added to the network together with $m_t$ incident edges; $m_t$ is a random variable assigned according to a Poisson distribution with expectation $\lambda > 0$.
	\item[{\it Preferential attachment}:] Each edge emanating from $v_{t+1}$ is connected with a node already in the network. Node selection can be considered as a series of $m_t$ independent trials, where at each trial the probability of selecting a node from $V_t$ with degree $k$ is
\begin{equation}
q_t(k) = \frac{r(k)}{\sum_{i=1}^{t} r(k_{i,t})},
\end{equation} 
where $k_{i,t}$ is the degree of node $v_i$ at step $t$. Define $s_{i,t}$ as the number of times node $v_i$ is chosen at step $t$. Then the entire selection procedure is equivalent to drawing a vector $(s_{1,t},s_{2,t}, \ldots, s_{t,t})$ from a multinomial distribution with probabilities $q_t(k_{1,1})$, \ldots, $q_t(k_{t,t})$ and sample size $m_t$. Equivalently, $s_{i,t}$ has Poisson distribution with expectation $\lambda q_t(k_{i,t})$ independently for $i=1,\ldots,t$.    
\end{description}

The PG model is determined by the choice of $r(k)$; we concentrate on two specifications and discuss their implications in the next section. Firstly, let 
\begin{equation} \label{eq:rk1}
	r(k)=k+a
\end{equation}
where the offset $a \geq 0$ is a constant. More generally, we define
\begin{equation} \label{eq:rk2}
	r(k) = k+a, \quad k\ge1, \quad \mbox{and} \quad r(0)=b
\end{equation} 
by taking $a \geq -1$ with extended domain, but in doing so  define a threshold parameter $b \geq 0$. Indeed, the latter formulation includes the former as a special case when we restrict $a=b \geq 0$, so that overall our model is specified by the parameter $\theta=(a,b,\lambda)$. 

The BA model can be explained as a reduction of our model by taking $a=b=0$, and by fixing $1 \leq m_t=m \leq t_0$ so that the number of edges added to the system at each step is a constant; the new edges are preferentially attached from the new node to {\it exactly} $m$ other nodes. Many structural constrains are implicit in the BA model. Indeed, at step $t$, a network with $t$ nodes must have $m(t-t_0)+|E_{t_0}|$ edges, none of which are multiple, whereas the number of edges for the PG model can take other values.  

A number of extensions of the BA model based upon generalizing $r(k)$ have been proposed. In particular, \cite{Krapivsky2} analyzed a version where the preferential attachment function is not linear in the degree $k$ of a node, but instead can be a power of the degree $k^\nu, \nu > 0$. They showed that for the scale-free property to hold, $r(k)$ must be asymptotically linear in $k$. In a subsequent work, \cite{Krapivsky} and~\cite{Dorogovtsev} independently went on to establish that adding the offset $a>-m$ as in~(\ref{eq:rk1}) does not violate this requirement, and derived the power-law exponent $\gamma=3+a/m$. Their result is analogous with our reported power-law exponent in~(\ref{eq:gamma1}) with $\lambda=m$ as seen in the next section. Furthermore, \cite{Krapivsky} investigated an attachment function similar to~(\ref{eq:rk2}) defined by $r(k)=k, k \geq2, r(1)=b, r(0)=0$. As they took $m \geq 1$ they did not need to be concerned with nodes of degree $k=0$. The power-law exponent they derived in this case is reminiscent of our result in~(\ref{eq:gamma2}). 

\section{The Degree Distribution of the Poisson Growth Model} \label{sec: degree distribution}

In this section, we discuss the degree distribution $p(k)$ for networks generated under the PG model. The main result is that the degree distribution follows a power-law
\begin{equation} \label{eq:pklaw}
 p(k) \sim k^{-\gamma},
\end{equation}
where $a_k \sim b_k$ indicates these two sequences are proportional to each other so that $a_k/b_k$ converges to a nonzero constant as $k\to\infty$. This result is an immediate consequence of the recursive formula
\begin{equation} \label{eq:pkrec}
 (k+a-1+\gamma) p(k) = (k+a-1) p(k-1)
\end{equation}
for sufficiently large $k$, and thus $p(k)\sim (k+a-1)^{-\gamma}\sim k^{-\gamma}$. The power-law exponent is
\begin{equation} \label{eq:gamma2}
 \gamma = 3+ \frac{a + (b-a)p(0)}{\lambda}
\end{equation}
for the preferential attachment function defined in~(\ref{eq:rk2}), and the exponent takes the range $\gamma>2$; the lower limit $\gamma\to2$ can be attained by letting $a=-1$, $b=0$, and $\lambda\to0$. This lower limit is in fact the limit for any form of $r(k)$ when $\lambda$ does not depend on $t$; $\gamma$ must be larger than 2 to ensure the mean degree $\sum_{k=0}^\infty k p(k) = 2\lambda$ converges.  For the special case~(\ref{eq:rk1}), the exponent becomes
\begin{equation} \label{eq:gamma1}
 \gamma = 3 + \frac{a}{\lambda},
\end{equation}
and the range is $\gamma\ge3$.

To make the argument precise, we have to note that $G_t$ is generated randomly and the degree distribution of $G_t$ also varies.  Let $n_t(k)$ be the number of nodes in $G_t$ with degree $k$. Since $\sum_{k=0}^\infty n_t(k)=t$, the observed degree distribution of $G_t$ is defined by $p_t(k)=n_t(k)/t$ for $k\ge0$.  For sufficiently large $t$, $p_t(k)$ may follow the power-law of (\ref{eq:pklaw}) as seen
below.

We consider a moderately large $k$ for the asymptotic argument as $t\to\infty$.  The maximum value of $k$ for consideration is $k \sim t^c$ for a given $t$ with a constant $c = 1/(\gamma+2+\epsilon)$ with any $\epsilon>0$. Then, the expectation of $p_t(k)$ can be expressed as
\begin{equation} \label{eq:eptk}
 \E(p_t(k)) \sim k^{-\gamma},
\end{equation}
which is the power-law we would like to show for $p_t(k)$. The variance of $p_t(k)$ will be shown as
\begin{equation} \label{eq:vptk}
 \V(p_t(k)) = O(k^{2+\epsilon} t^{-1}),
\end{equation}
indicating the variance reduces by the factor $1/t$. Note that (\ref{eq:vptk}) is not a tight upper bound, and the variance can be much smaller. See the Appendix for the proof of (\ref{eq:eptk}) and (\ref{eq:vptk}).  Let $0<d<1/(2\gamma+2+\epsilon)$, and consider $k = O(t^d)$, which is even smaller than $t^c$. Then,
\begin{equation} \label{eq:pkvariation}
\frac{\sqrt{\V(p_t(k))}}{\E(p_t(k))}
= O(k^{\gamma+1+\epsilon/2}t^{-1/2})
= O(t^\alpha)
\end{equation}
with $\alpha = d(\gamma+1+\epsilon/2)-1/2<0$, and thus the limiting distribution $\lim_{t\to\infty} p_t(k)=p(k)$ follows the power-law of (\ref{eq:pklaw}). By taking $\epsilon\to0$, the power-law of $p_t(k)$ is shown up to $k\sim t^d$ with $d<1/(2\gamma+2)$.

It remains to give an expression for $p(0)$ in (\ref{eq:gamma2}). We will show in the Appendix that $p(0)$ is a solution of the quadratic equation
\begin{equation} \label{eq:p0quad}
   (b-a) x^2 + 
(2\lambda +a + \lambda b - (b-a)e^{-\lambda}) x
-(2\lambda+a) e^{-\lambda} = 0.
\end{equation}
For $a\neq b$, one of the solutions 
\begin{eqnarray}
 p(0) &=& \frac{1}{2(b-a)}\Bigl[
\left\{
(2\lambda +a + \lambda b - (b-a)e^{-\lambda})^2
+4(b-a)(2\lambda+a) e^{-\lambda}
\right\}^{1/2} \nonumber \\
&&\hspace{5em} - (2\lambda +a + \lambda b - (b-a)e^{-\lambda})
\Bigr]  \label{eq:pzero}
\end{eqnarray}
is the unique stable solution with $0<p(0)<1$; this can be checked by looking at the sign of $p_{t+1}(0) - p_t(0)$ in the neighborhood of $p(0)$. 

\section{Simulation Study} \label{sec: simulation}

A small simulation study was conducted to support our theoretical claims of Section~\ref{sec: degree distribution}. Specifically, we wish to confirm via simulation that the degree distribution $p(k)$ of~(\ref{eq:pklaw}) as well as its expected value $\E(p_t(k))$ as in~(\ref{eq:eptk}) follow a power-law with $\gamma$ as in~(\ref{eq:gamma2}). To that end we generated networks under the PG model for a variety of parameter settings. For each specification of $\theta$ we generated $n_{sim}=10^4$ networks of size $N=5000$, each from a seed network of a pair of connected nodes. We included the BA model, generated under analogous conditions, so as to demonstrate the soundness of our results which are summarized in Table~\ref{tab:1}.

\begin{figure}
\begin{center}
\includegraphics[width=12cm]{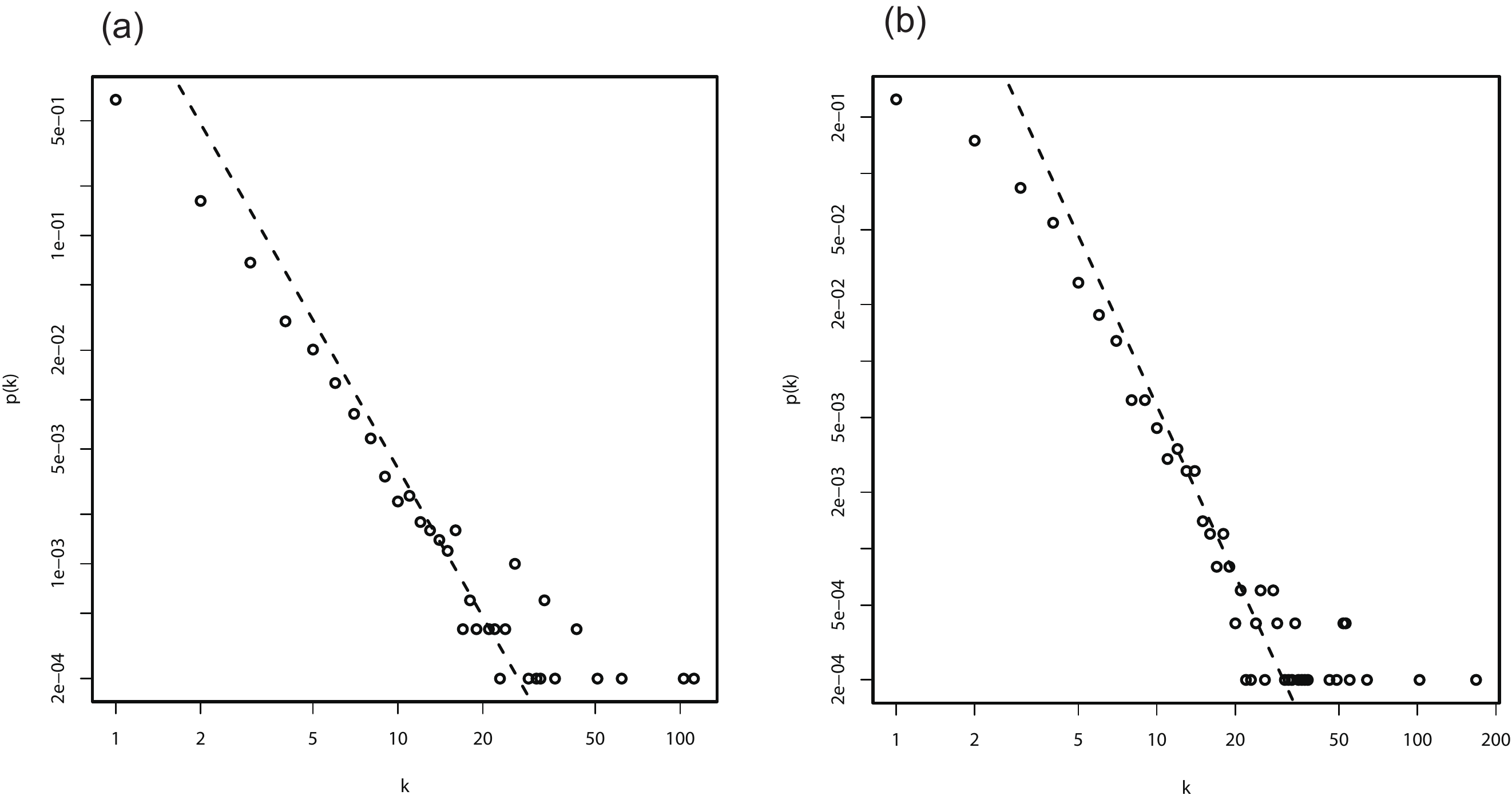}  
\caption{Degree distribution $p(k)$ of a typical network plotted on a log-log
 scale with the power-law line using estimated exponent $\hat\gamma$.
(a)~Generated under the BA model; $\hat{\gamma}=3.03$.
(b)~Generated under the PG model with $\theta=(0,0,1)$; $\hat{\gamma}=3.01$.}
\label{fig:pk}       
\end{center}
\end{figure}

In point of fact, estimating $\gamma$ from a network can be quite tricky and it has been the subject of some attention in the literature; see~\cite{Goldstein}. We sided with the maximum likelihood (ML) approach described by~\cite{Newman2}. In this methodology, the ML estimate of $\gamma$ for a particular network is given by
\begin{eqnarray*}
	\hat{\gamma}=1+ \left(\sum_{k\ge k_{min}} n(k)\right) \cdot \left(
			 \sum_{k\ge k_{min}} n(k)\log \frac{k}{k_{min}} \right)^{-1}
\end{eqnarray*}
where $n(k)$ is the number of nodes with degree $k$, and $k_{min}$ is the minimum degree after which the power-law behavior holds. \cite{Bauke} studied selecting a value for $k_{min}$ by using a $\chi^2$ goodness of fit test over a range of $k_{min}$; however, we shied away from this level of scrutiny as we found that taking $k_{min}=10$ was reasonable for our examples. This methodology is illustrated in Figure~\ref{fig:pk} (a) and (b) where we plot the degree distribution with $\hat\gamma$ for a typical network generated by the BA and PG model, respectively.

Returning to Table~\ref{tab:1}, in each case, we confirm (\ref{eq:pklaw}) has power-law exponent as predicted by~(\ref{eq:gamma2}) and (\ref{eq:pzero}).  We computed $\hat{\gamma}$ for each network, and calculated the mean and standard deviation of $\hat\gamma$ values for $n_{sim}$ networks. We observe that the mean $\hat\gamma$ agrees well with the predicted $\gamma$, and the variation of $\hat\gamma$ is relatively small as suggested by (\ref{eq:pkvariation}).

\begin{figure}
\vspace{2cm}
\begin{center}
\includegraphics[width=12cm]{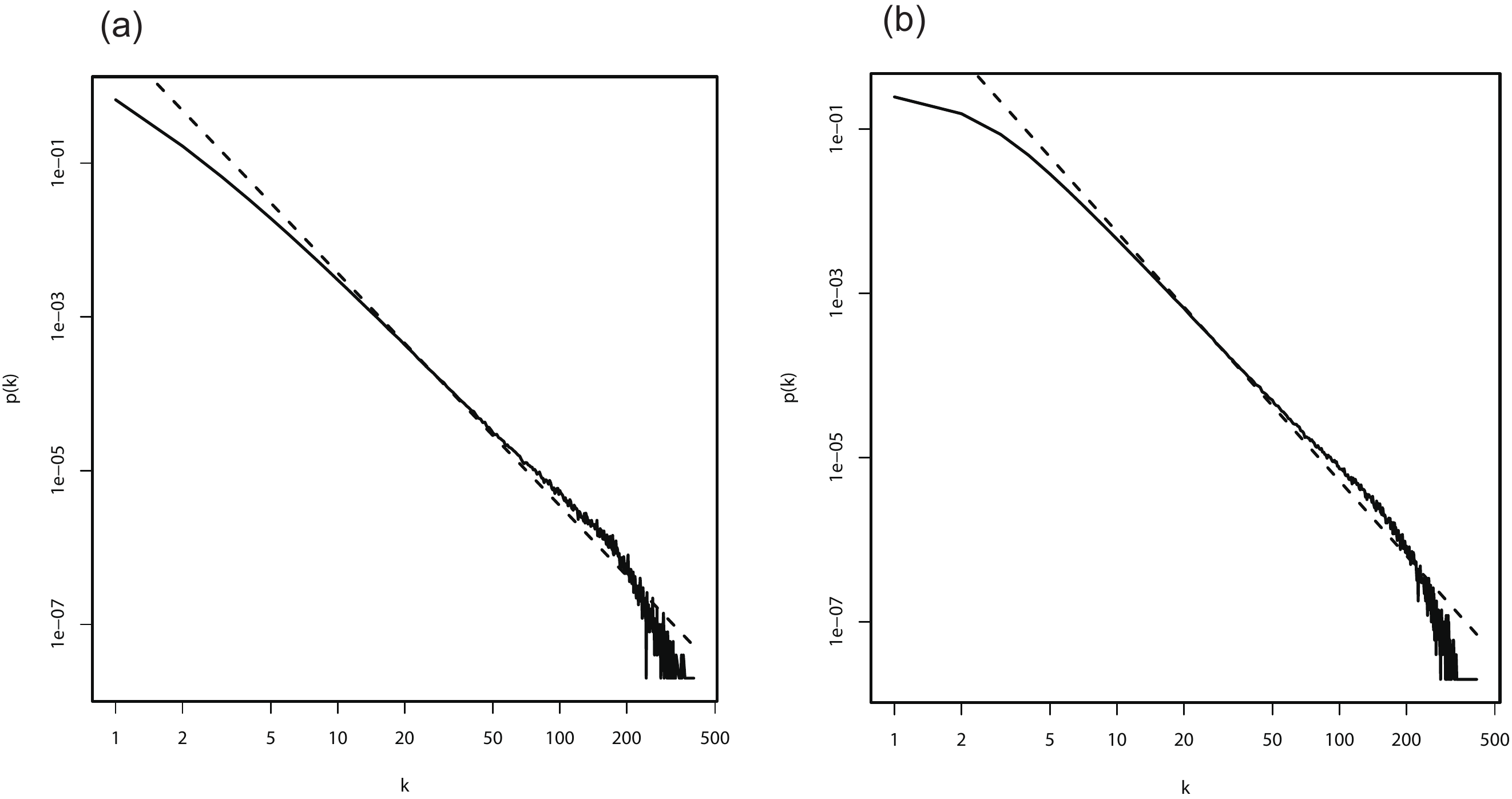}  
\caption{Avrage degree distribution $\E(p(k))$ of the simulation with the power-law
 line using estimated exponent $\hat\gamma_{avg}$. Ploted for (a)~the BA
 model and for (b)~the PG model with $\theta=(0,0,1)$, where
 $\hat\gamma_{avg}=3.03$ for both cases.}
\label{fig:Epk}       
\end{center}
\end{figure}

In addition, to show that the same holds for~(\ref{eq:eptk}), in each case we computed the average degree distribution of the $n_{sim}$ networks as an estimate of $\E(p(k))$. Then we estimated the degree exponent $\hat{\gamma}_{avg}$ as seen in the table and Figure~\ref{fig:Epk}. Again, the simulated results match well with theory.

%\begin{quotation}
%----- Table 1, and Figures 1, 2 are inserted around here -----
%\end{quotation}

\section{Discussion} \label{sec: discussion}

The PG model has a special place in the class of preferential attachment models. It has a tunable power-law exponent and a simple implementation, yet it can generate any network. In contrast, the BA model and its generalizations described in Section~\ref{sec: PG model} have serious restrictions on the types of networks that can be generated because $m$ is held constant. For example, at step $t$ an instantiation of the BA model will consist of a $t$ node network with the number of edges equal to exactly $m(t-t_0)$, plus the number of edges in the seed network. The simple design of our model makes computing the probability of a network straightforward. This in combination with its modeling potential gives rise to several useful applications in Bayesian inference.

In explicit terms, let $G=(V,E)$ be a network with $N=|V|$ nodes where $V=\{v'_1,\ldots,v'_N\}$. Furthermore, let $G_N=(V_N,E_N)$ be a network generated under PG model after step $N-1$ so that $V_N=\{v_1,\ldots,v_N\}$, where the seed network consists of a single node.  The association between $V$ and $V_N$ is defined by a permutation $\sigma=(\sigma_1,\ldots,\sigma_N)$ so that $v_i = v'_{\sigma_i}$. Given $G$, once we specify $\sigma$, then it is straightforward to compute $k_{i,t}$, $s_{i,t}$ for $i=1,\ldots,t$; $t=1,\ldots,N-1$. Then the probability of $G$ given $\theta=(a,b,\lambda)$ and $\sigma$ is
\[
 \P(G | \theta,\sigma) = \prod_{t=1}^{N-1} \left(\prod_{i=1}^t
       e^{-\lambda q_t(k_{i,t})} \frac{(\lambda
       q_t(k_{i,t}))^{s_{i,t}}}{s_{i,t}!}
\right).
\]

One application is when $G$ is known and we wish to estimate $\theta$. This can be done by assigning a prior $\pi(\theta)$ for $\theta$ and the uniform prior on $\sigma$. The posterior probability of $(\theta,\sigma)$ given $G$ is 
\[
 \pi(\theta,\sigma | G) \propto \P(G | \theta,\sigma) \pi(\theta).
\]
Using MCMC to produce a chain of values for $(\theta,\sigma)$, the posterior $\pi(\theta|G)$ is simply obtained from the histogram of $\theta$ in the chain. Moreover, this procedure can be used for model comparison, if we have several models for generating the network.

Another application is when we wish to make inference about $G$ from data $D$ with likelihood function $P(D|G)$. The posterior probability of $(G,\theta,\sigma)$ given $D$ is
\[
 \pi(G,\theta,\sigma | D)
\propto P(D|G) \P(G | \theta,\sigma) \pi(\theta).
\]
Then the posterior $\pi(G|D)$ is simply obtained from the frequency of $G$ in the chain. Indeed, we used this approach for inferring a gene network from microarray data in~\cite{Sheridan}.

Recall that the PG model produces networks with multiple edges. In practice, we often want to restrict our interest to networks without multiple edges. As an approximation, we could apply the formula for $P(G|\theta,\sigma)$ just as well in this case.  Alternatively, we propose a slight modification to our model where we generate $m_t$ edges at step $t$ according to a binomial distribution with parameter $p=\lambda/t$ and sample size $t$. In this formulation the seed network must be selected such that $\lambda \leq t_0$, otherwise $p>1$ may occur. Then by sampling nodes without replacement, multiple edges are avoided. In our simulation (results not included) we found that these modifications do not change the power-law.

Finally, though we made specific choices for $r(k)$ in our arguments, the PG model can be generalized to a wider class of preferential attachment functions. For instance, \cite{Dorogovtsev2} investigated {\em accelerated growth} models where $m_t$ increases as the network grows. It should be possible to incorporate accelerated growth into PG model by gradually increasing the value of $\lambda$ over time. Another line of generalizations of the PG model is via the inclusion of local events. 

\section*{Appendix: Proofs}

\bigskip
\noindent
{\large\it The expected value of $p_t(k)$}
\medskip

Here we give the proof  of (\ref{eq:eptk}). We assume that the functional form of $r(k)$ is (\ref{eq:rk1}), and a modification to handle (\ref{eq:rk2}) is mentioned at the bottom. 

Let $I(A)$ denote the indicator function of the event $A$, so $I(A)=1$ if $A$ is true and $I(A)=0$ if $A$ is false. We use the notation $\P(\cdot)$, $\E(\cdot)$ and $\V(\cdot)$ for the probability, expectation and the variance, and also $\P(\cdot|A)$, $\E(\cdot |A)$ and $\V(\cdot |A)$ for those given a condition $A$. By noting
\[
  n_{t+1}(k) = \sum_{i=1}^t I(k_{i,t}+s_{i,t} = k) + I(m_t=k),
\]
the conditional expectation of $n_{t+1}(k)$ given $G_t$ is
\begin{eqnarray}
\E( n_{t+1}(k)  | G_t) 
&=& 
\sum_{i=1}^t \P(k_{i,t}+s_{i,t} = k| G_t) + 
\P(m_{t}=k | G_t) \nonumber\\
&=&
\sum_{i=1}^t e^{-\lambda q_t(k_{i,t})} \frac{(\lambda
 q_t(k_{i,t}))^{k-k_{i,t}}}{(k-k_{i,t})!} 
+ e^{-\lambda} \frac{\lambda^k}{k!} \nonumber\\
&=&
\sum_{s=0}^k n_t(k-s) e^{-\lambda q_t(k-s)}\frac{(\lambda
       q_t(k-s))^s}{s!} + e^{-\lambda} \frac{\lambda^k}{k!}.
 \label{eq:enk1}
\end{eqnarray}

The last term $e^{-\lambda} \lambda^k/k! \sim (e\lambda/k)^k$ can be
ignored for a large $k$, since it is exponentially small as $k$ grows.
We examine the terms in the summation over $s=0,1,\ldots,k$ for
$k=O(t^c)$ as $t\to\infty$. For a fixed $s$, $q_t(k-s) \sim k/t $ for a
linear preferential attachment model. More specifically,
for $r(k)=k+a$, $k\ge0$,
\[
 q_t(k-s)=\frac{r(k-s)}{\sum_{i=1}^t r(k_{i,t})}
=
\frac{k-s+a}{t(2\lambda + a)}(1+O(t^{-1/2})),
\]
because the mean degree of $G_t$ is
\[
 \frac{1}{t}\sum_{i=1}^t k_{i,t} =  \frac{2}{t}
\left(
|E_{t_0}|+
\sum_{t'=t_0}^{t-1} m_{t'}\right) =
       2 \lambda + O(t^{-1/2}),
\]
and the denominator of $q_t(k)$ is
\begin{equation} \label{eq:sumrk}
 \sum_{i=1}^t r(k_{i,t}) =  \sum_{i=1}^t (k_{i,t}+a) = t(2\lambda + a +
 O(t^{-1/2})).
\end{equation}
Thus the sum in (\ref{eq:enk1}) over $s=0,1$ becomes
\[
 n_t(k) \left(
1 - \frac{\lambda (k+a)}{(2\lambda+a)t}
+O(k t^{-3/2})
\right)
+ n_t(k-1)\left(
\frac{\lambda (k+a-1)}{(2\lambda+a)t} +
O(k t^{-3/2})
\right).
\]
For $s\ge 2$, each term is $\sim n_t(k-s) (k/t)^s $. By noting
$\sum_{s=2}^k n_t(k-s)\le t$, the sum over
$s=2,\ldots,k$ becomes $O(k^2 t^{-1})$.

Next, we take the expectation of (\ref{eq:enk1}) with respect to $G_t$
to obtain the unconditional expectation $\E(n_{t+1}(k))$, and replace
$n_t(k)=t p_t(k)$. Using the results of the previous paragraph, we get
\begin{eqnarray}
  \E(p_{t+1}(k)) =  \E(p_t(k)) 
-\frac{\lambda}{(2\lambda+a)t}
\biggl(
(k'+\gamma+O(k t^{-1/2}))\E(p_t(k)) \nonumber\\
-(k'+O(k t^{-1/2}))\E(p_t(k-1))
+O(k^2 t^{-1})
\biggr) \label{eq:eptupdate}
\end{eqnarray}
with $k'=k+a-1$ and the $\gamma$ of (\ref{eq:gamma1}).  Let us assume
$\E(p_t(k-1))\sim (k-1)^{-\gamma}$, and remember $c < 1/(\gamma+2)$.
By taking the limit $t\to\infty$ and equating
       $\E(p_{t+1}(k))=\E(p_t(k))$,
 we get
\[
(k' +\gamma +  o(1)) \E(p_t(k))=(k'+  o(1))\E(p_t(k-1)). 
\]
So that, for sufficiently large $t$,
\[
\E(p_t(k)) \sim k^{-\gamma}
\]
also holds for $k$. Since $\E(p_{t}(k))=O(1)$ for a fixed $k$, the
power-law holds for any $k$ by induction up to $k\sim t^c$.

For $r(k)$ of (\ref{eq:rk2}), the preferential attachment is modified to
\[
 r(k) = k + a +  (b-a) I(k=0), \quad k\ge0.
\]
This changes the the denominator of $q_t(k)$ in (\ref{eq:sumrk}) to
\begin{equation} \label{eq:sumrk2}
  \sum_{i=1}^t r(k_{i,t}) = 
 t\left(2\lambda + a + (b-a)p_t(0) + O(t^{-1/2})\right),
\end{equation}
and thus $2\lambda+a$ in the updating formula (\ref{eq:eptupdate}) is replaced with $2\lambda +a + (b-a)p(0) $, leading to (\ref{eq:gamma2}). Note that $p_t(0)=p(0)+O(t^{-1/2})$ from (\ref{eq:vptk}) shown in the next section.

\bigskip
\noindent
{\large\it The variance of $p_t(k)$}
\medskip

Here we give the proof of (\ref{eq:vptk}) by working on $\V(n_t(k))=t^2
\V(p_t(k))$.  Although $r(k)$ of (\ref{eq:rk1}) is again assumed, the
argument is basically the same for (\ref{eq:rk2}). By noting the
identity
\begin{equation} \label{eq:vntidentity}
\V(n_{t+1}(k)) = \E( \V( n_{t+1}(k)  | G_t))
+
 \V( \E( n_{t+1}(k)  | G_t) ),
\end{equation}
we evaluate the two terms on the right hand side.

The conditional variance of $n_{t+1}(k)$ given $G_t$ is evaluated rather
similarly as the conditional expectation of (\ref{eq:enk1}). By noting
$\V(I(A))=\P(A)-\P(A)^2$, $ \V(n_{t+1}(k)|G_t)$ is expressed for $k
=O(t^c)$ as
\[
\sum_{s=0}^k n_t(k-s) \left\{e^{-\lambda q_t(k-s)}\frac{(\lambda
       q_t(k-s))^s}{s!}
-\left(e^{-\lambda q_t(k-s)}\frac{(\lambda
       q_t(k-s))^s}{s!}\right)^2
\right\}
\]
\begin{equation} \label{eq:vn1}
 \approx n_t(k) \frac{\lambda(k+a)}{(2\lambda+a)t}
+ n_t(k-1) \frac{\lambda(k+a-1)}{(2\lambda+a)t},
\end{equation}
where terms from $I(m_t = k)$ are ignored for a large $k$.
Thus, the first term in (\ref{eq:vntidentity}) is
\[
\E(\V(n_{t+1}(k)|G_t)) = O(k^{-\gamma+1}).
\]

On the other hand, the second term in (\ref{eq:vntidentity}) is
evaluated by considering the variance of (\ref{eq:enk1}) as
\begin{eqnarray*}
\V( \E(n_{t+1}(k)|G_t) ) &\le&
 \V(n_t(k)) \left(  1 -  \frac{2\lambda(k+a)}{(2\lambda+a)t}
+O(k t^{-3/2})
\right)\\
&&\hspace{-5em}+2\sqrt{\V(n_t(k))}\sqrt{\V(n_t(k-1))}
\left(
\frac{\lambda(k+a-1)}{(2\lambda+a)t} + O(k t^{-3/2})
\right)\\
&&\hspace{-5em}+\V(n_t(k-1))O(k^2 t^{-2}) + \sqrt{\V(n_t(k))} O(k^2 t^{-1}) + O(k^4 t^{-2}).
\end{eqnarray*}

We substitute these two expressions for those in (\ref{eq:vntidentity}).
We will show, by induction, that
\begin{equation} \label{eq:vntkv}
 \V( n_t(k) ) < A k^{2+\epsilon} t
\end{equation}
holds for all $(t,k)$ with $k=O(t^c)$ using some constant $A$. Let us
assume that (\ref{eq:vntkv}) holds for $(t,k)$ and $(t,k-1)$. 
By taking a sufficiently large $A$, we have
\begin{equation} \label{eq:vntupdate}
 \V(n_{t+1}(k)) \le A k^{2+\epsilon} (t- (2\lambda+a)^{-1}) +
 o(k^{1+\epsilon/2})
< A k^{2+\epsilon}(t+1),
\end{equation}
implying that (\ref{eq:vntkv}) also holds for $(t+1,k)$. 

On the other hand, for any random variable $0\le n \le t$ with its
expectation $\E(n)$ fixed, the largest possible variance $O(t) \E(n)$ is
attained if the probability concentrates on the extreme values 0 and
$t$. Applying this upper bound to $n_t(k)$ with $k\sim t^c$, we obtain
$\V(n_t(k))/t = O(\E(n_t(k))) = O(k^{-\gamma}
t)=O(k^{2+\epsilon})$, implying that (\ref{eq:vntkv}) holds for any
$(t,k)$ with $k\sim t^c$.

For induction with respect to $k$, we only have to show 
\begin{equation} \label{eq:vt2}
 \V(n_t(k)) < v(k) t
\end{equation}
for a sufficiently large $k$ so that terms from $I(m_t =
k)$ in (\ref{eq:vn1}) can be ignored. $v(k)$ is an arbitrary constant
depending on $k$.  We start from $k=0$.  First note that
\[
 n_{t+1}(0) = \sum_{i=1}^t I(k_{i,t}=0 \cap s_{i,t}=0) + I(m_t = 0).
\]
Thus $\E(n_{t+1}(0)|G_t) = n_t(0) e^{-\lambda q_t(0)} + e^{-\lambda}$, and so
\[
 \V(\E(n_{t+1}(0)|G_t) ) = \V(n_t(0))\left(
1 - \frac{2\lambda a}{(2\lambda+a)t} + O(t^{-3/2})
\right).
\]
On the other hand, $ \V(n_{t+1}(0)|G_t)$ is expressed as
\[
  n_t(0)(
e^{-\lambda q_t(0)} - e^{-2\lambda q_t(0)}
) +
 e^{-\lambda} - e^{-2\lambda}
 + 2n_t(0)  (1-e^{-\lambda q_t(0)}) e^{-\lambda}.
\]
By substituting these two expressions for those in
(\ref{eq:vntidentity}), we observe that the increase of the variance,
i.e., $\V(n_{t+1}(0))- \V(n_t(0))$ is bounded by a constant, and we have
$\V(n_t(0)) = O(t)$.

Let us assume (\ref{eq:vt2}) holds up to $k-1$. Then $\V(n_{t+1}(k))$
can be expressed quite similarly as (\ref{eq:vntupdate}), but
$\E(\V(n_{t+1}(k)|G_t))$ includes additional terms from $I(m_t=k)$;
$\V(I(m_t=k))=O(1)$ and $\E(\sum_{i=1}^t {\rm Cov}(I(k_{i,t}+
s_{i,t}=k), I(m_t=k) | G_t))$. For $k_{i,t}=k$, the covariance term $\le
\P(m_t=k)(1-\P(s_{i,t}=0|G_t))=O(t^{-1})$, and for $k_{i,t}=k-s$ with
$s\ge1$, the covariance term $\le \P(s_{i,t}=s)(1-\P(m_t = k)) =
O(t^{-s})$. Thus, by taking the sum over $i=1,\ldots,t$, it becomes $O(t\cdot
t^{-1})=O(1)$. Therefore, $\V(n_{t+1}(k))- \V(n_t(k))$ is bounded by a
constant, and (\ref{eq:vt2}) holds for $k$. By induction, (\ref{eq:vt2})
holds for any $k$.

\bigskip
\noindent
{\large\it The equation of $p(0)$}
\medskip

Here we derive (\ref{eq:p0quad}) for the $r(x)$ of (\ref{eq:rk2}).  By taking the expectation of $ \E(n_{t+1}(0)|G_t) = n_t(0) e^{-\lambda q_t(0)} + e^{-\lambda}$ with respect to $G_t$, and using (\ref{eq:sumrk2}), we get
\[
 \E(n_{t+1}(0)) = \E(n_t(0)) \left(
1 - \frac{\lambda b}{(2\lambda + a + (b-a)p(0))t}
+ O(t^{-3/2})
\right) + e^{-\lambda}.
\]
By substituting $n_t(0)=t p_t(0)$ and taking the limit $t\to\infty$, we get a formula for $f(x)=(t+1)(\E(p_{t+1}(0))-\E(p_t(0)))$ as a function of $x=p(0)$ 
\[
 f(x) = -x\left( 1+ \frac{\lambda b}{2\lambda+a +(b-a)x }\right) +
 e^{-\lambda}.
\]
The quadratic equation (\ref{eq:p0quad}) is obtained by letting $f(x)=0$. In addition, the condition $d f(x)/d x <0$ was checked for the stable solution.

\clearpage

\begin{table}
\caption{Summary of estimated power-law exponents from simulated
 networks. The last column is theoretically predicted $\gamma$.
}
\label{tab:1}       
\begin{tabular}{llllll}
\hline\noalign{\smallskip}
Model & 	Parameters 				&     Mean $k$
 &  Mean $\hat\gamma \ \pm $ s.d. & 	$\hat{\gamma}_{avg}$
 & $\gamma$ 		 \\
\noalign{\smallskip}\hline\noalign{\smallskip}
BA & $m=1$ 					& 	2.0 				&  	$3.03 \pm 0.15$ 		& 	3.03					&	 3	 				\\
PG & $\theta=(0,0,1)$	 		& 	2.0 				& 	$3.03 \pm 0.12$ 		&	3.03					&	 3			 		\\
PG & $\theta=(-0.9,0.1,1)$	 	& 	2.0 				& 	$2.54 \pm 0.10$ 		&	2.51					&	 2.44 				\\
PG & $\theta=(-0.9,0.1,3)$	 	& 	6.0 				& 	$2.86 \pm 0.05$ 		&	$2.86$					&	 2.72 				\\
PG & $\theta=(0.5,0.5,3)$	 		& 	6.0 				& 	$3.15 \pm 0.05$ 		&	3.15					&	 3.17		 			\\	
\noalign{\smallskip}\hline
\end{tabular}
\end{table}

\clearpage

\end{document}